\documentclass[useAMS,usenatbib]{mn2e}
\usepackage[]{natbib,amsmath,amssymb,times}
\bibpunct{(}{)}{;}{a}{}{,}

\usepackage[utf8]{inputenc}
\usepackage{eurosym,graphicx,hyperref,txfonts,epsfig,latexsym,ulem}

\renewcommand{\d}{\mathrm{d}}
\newcommand{\e}{\mathrm{e}}
\newcommand{\ii}{\mathrm{i}}
\newcommand{\bea}{\begin{eqnarray}}
\newcommand{\eea}{\end{eqnarray}}
\newcommand{\be}{\begin{equation}}
\newcommand{\ee}{\end{equation}}

\newcommand{\vc}[1]{\mbox{\boldmath $#1$}}

\title[spin-3 flexion]
{Estimation of halo ellipticity using spin-3 flexion}
\author[]%
{Xinzhong Er$^1$\thanks{xer@nao.cas.cn},
  Matthias Bartelmann$^2$\thanks{bartelmann@uni-heidelberg.de},
\\
$^1$National Astronomical Observatories, Chinese Academy of Sciences,
Beijing 100012, China\\
$^2$Zentrum f\"ur Astronomie, ITA, Universit\"at Heidelberg,
Albert-Ueberle-Str.\ 2, 69120 Heidelberg, Germany\\
}%

\date{Accepted ---; received ---; in original form ---}
\pubyear{2012}

\begin{document}

\maketitle
\begin{abstract}
  Estimating the ellipticity of dark matter haloes at the galaxy or galaxy-cluster scale can provide important constraints on the formation of galaxies or clusters, as well as on the nature of dark matter. We show in this paper that the spin-3 gravitational flexion can add useful information on the ellipticity of lensing haloes. We introduce a general formalism to decompose fields with arbitrary spin into radial and tangential components. The ratio of the tangential and radial flexion components directly estimates the lens ellipticity. We point out that any centroid offset will significantly bias our estimate, which on the other hand can be used to determine the centre of the lens halo.
\end{abstract}

\section{Introduction}

The structure of dark-matter haloes hosting clusters and galaxies is of substantial interests for cosmology. Halo properties consistently predicted by $N$-body simulations of structure formation in Cold Dark Matter (CDM) include a highly non-spherical structure, which is well fitted by a triaxial body with a characteristic density profile \citep{2002ApJ...574..538J, 2004IAUS..220..421S}. This property is generic for the nature of the dark matter as well as for the formation of galaxies and clusters \citep{2002sgdh.conf..109B, 2006MNRAS.367.1781A, 2010MNRAS.404.1137B, 2011MNRAS.416.1377V,2010ApJ...718..762W, 2011MNRAS.413.1973W}. For example, major mergers of dark-matter haloes may play an important role in forming their shapes \citep{1993ApJ...418..544V}. Testing halo ellipticities for their agreement with the cold-dark matter prediction may thus provide a potentially powerful constraint on the standard scenario for structure formation from galaxies to galaxy clusters.

Arguably the most promising tool to study the mass distribution of galaxy and cluster haloes is gravitational lensing \citep{2001PhR...340..291B}, as it probes the mass distribution independent of the nature of the matter (luminous or dark) or its dynamical state. Weak gravitational lensing is routinely being used for cluster mass reconstructions \citep[e.g.][]{2006ApJ...648L.109C, 2008ApJ...681..187B} and for measuring the ellipticity of dark-matter haloes \citep{2006MNRAS.370.1008M,2007MNRAS.380..149C, 2010ApJ...721..124D}. For example, a mean ellipticity of $0.46$ has been obtained for a sample of $25$ galaxy-cluster haloes based on Subaru data \citep{2010MNRAS.405.2215O}, roughly consistent with the CDM prediction.

Hitherto, the shear has mainly been used for weak-lensing studies, and the gravitational magnification in a few cases. Both are determined by the projected gravitational tidal field, i.e.~by linear combinations of second-order derivatives of the effective lensing potential. Measurements of third-order derivatives are now gradually coming within reach. They can conveniently be combined into quantities called the first and the second flexion, or $\mathcal{F}$ and $\mathcal{G}$ flexion, respectively. More on their properties will follow below.

The two flexions can be introduced as derivatives of either the surface mass density or the gravitational shear. They respond to smaller-scale variations in the projected mass distribution than the shear and the surface-mass density \citep{2002ApJ...564...65G, 2005ApJ...619..741G, 2006MNRAS.365..414B}. Several techniques have been proposed to measure flexion \citep[e.g.][]{2007ApJ...660..995O, 2011MNRAS.412.2665V, 2011ApJ...736...43C}, but an unambiguous measurement has yet to be demonstrated. It has recently been shown that a reliable estimate of the flexions requires an accurate model for the shear, which must be taken into account in any unbiased flexion estimator \citep{2012MNRAS.419.2215V}.

The intrinsic noise of measured flexion components is still poorly understood \citep{2007ApJ...660.1003G}. However, it has been suggested that flexion can significantly improve the estimation of halo ellipticity \citep{2012MNRAS.421.1443E}. In particular, \citet{2011A&A...528A..52E} have introduced the ratio of the tangential and radial components of the $\mathcal{F}$ flexion to estimate the halo ellipticity. Since this approach is independent of the lens strength, and thus also independent of the mass of the lens, its redshift and even the halo-centric distance, one should be able to use this ratio of $\mathcal{F}$ flexion components to estimate the halo ellipticity as a function of radius \citep{2011MNRAS.417.2197E}.

Here, we extend this earlier study in three ways, motivated by the following considerations: First, the $\mathcal{F}$ flexion is a quantity with the transformation properties of a vector. It is thus straightforward to define its radial and tangential components. The $\mathcal{G}$ flexion has a three-fold rotational symmetry, which renders its projection on tangential or radial directions less transparent. We propose here a general formalism for decomposing lensing quantities of arbitrary order and spin weight into components of different orientations, which may also help clarifying the properties of the $\mathcal{F}$ and $\mathcal{G}$ flexions. Second, the $\mathcal{F}$ flexion leads to a centroid shift in images which is probably hard to measure reliably. The $\mathcal{G}$ flexion with its unique, triangular distortion pattern should be considerably easier to access. It is therefore important to exploit in particular the signal of the $\mathcal{G}$ flexion, if flexion can be measured at all. Third, since the signal is expected to be weak, the combination of all accessible signals is mandatory, calling for the $\mathcal{G}$ flexion to be included in measurements of halo shapes.

In Sect.~2, we introduce our formalism for decomposing spin-weighted fields of arbitrary spin into tangential and radial components, and apply it specifically to the gravitational flexion. We show the results of numerical tests in Sect.~3. We discuss these results and summarise our conclusions in Sect.~4. Where necessary throughout this paper, we adopt a $\Lambda$CDM standard cosmological model with $\Omega_{\Lambda}=0.75$, $\Omega_{\rm m}=0.25$, and a Hubble constant of $H_0 = 73$ km\,s$^{-1}$\,Mpc$^{-1}$.

\section{Formalism}

\subsection{Preliminaries}

The fundamentals of gravitational lensing can be found in \citet{2001PhR...340..291B, 2006MNRAS.365..414B}. For its elegance and brevity, we shall use the complex notation for shear and flexion. The thin-lens approximation is adopted, implying that the lensing mass distribution can be projected onto the lens plane perpendicular to the line-of-sight. We introduce angular coordinates $\vec\theta$ with respect to the line-of-sight. The lensing convergence, i.e., the dimensionless projected surface-mass density, can be written as $\kappa(\vec\theta) = \Sigma(\vec\theta)/\Sigma_{\rm cr}$, where $\Sigma(\vec\theta)$ is the projected surface-mass density and $\Sigma_{\rm cr}$ is its critical value
\be
  \Sigma_{\rm cr} = \frac{c^2}{4\pi G} \frac{D_{\rm s}}{D_{\rm d} D_{\rm ds}}\;,
\label{eq:1a}
\ee
depending on the angular-diameter distances $D_{\rm s}$, $D_{\rm d}$ and $D_{\rm ds}$ from the observer to the source, the observer to the lens, and the lens to the source, respectively. All lensing quantities can be derived from the effective lensing potential $\psi$,
\be
  \psi(\vc\theta) = {1\over \pi}\int_{{\cal R}^2} \d^2\theta'\kappa(\vc\theta')
  {\rm ln}|\vc\theta-\vc\theta'|\;.
\ee
To lowest order, image distortions caused by gravitational lensing are described by the complex shear
\be
  \gamma \equiv \gamma_1 + \ii \gamma_2
= \frac{1}{2}\left(\partial_1^2\psi-\partial_2^2\psi\right) + {\rm i}\partial_1\partial_2\psi\;.
\label{eq:1b}
\ee
The shear transforms a hypothetical round source into an elliptical image. The $\mathcal{F}$ and $\mathcal{G}$ flexions can be introduced as the complex derivatives
\bea
  \mathcal{F} &=& \left(\partial_1 + \mathrm{i}\partial_2\right)\kappa\;,\nonumber\\
  \mathcal{G} &=& \left(\partial_1 + \mathrm{i}\partial_2\right)\gamma\;.
\label{eq:1c}
\eea
The flexions are thus combinations of third-order derivatives of the effective lensing potential $\psi$. We shall denote their real and imaginary parts by $(\mathcal{F, G})_1$ and $(\mathcal{F, G})_2$, respectively. In terms of the lensing potential, we have
\begin{equation}
  \mathcal{F} \equiv {\cal F}_1 + \ii {\cal F}_2
= \frac{1}{2}\left(\partial_1^3\psi + \partial_1\partial_2^2\psi \right)+\frac{\ii}{2}\left(\partial_1^2\partial_2\psi + \partial_2^3\psi\right)
\label{eq:1d}
\end{equation}
and
\begin{equation}
  \mathcal{G} \equiv {\cal G}_1 + \ii {\cal G}_2
= \frac{1}{2}\left(\partial_1^3\psi - 3\partial_1\partial_2^2\psi \right)+\frac{\ii}{2}\left(3\partial_1^2\partial_2\psi-\partial_2^3\psi\right)\;.
\label{eq:1e}
\end{equation}

\subsection{Spin decomposition of tensors on the sphere}

We proceed by introducing a general formalism for decomposing arbitrary lensing distortions, or more generally fields on the sphere with arbitrary spin, into fields of defined rotational symmetry and projecting them onto arbitrary directions on the sky. For reference, we recommend Appendix~A of \citet{2005PhRvD..72b3516C} and the literature cited therein. Let $e_{1, 2}$ be the conventional coordinate basis vectors on the sphere,
\begin{equation}
  e_1 = \partial_\theta\;,\quad e_2 = \sin^{-1}\theta\partial_\phi\;,
\label{eq:2}
\end{equation}
and $\theta^{1, 2}$ their dual basis vectors,
\begin{equation}
  \theta^1 = \d\theta\;,\quad\theta^2 = \sin\theta\d\phi\;.
\label{eq:2a}
\end{equation}
We introduce the conventional helicity basis $e_\pm$ and its dual basis $\theta^\pm$,
\begin{equation}
  e_\pm = \frac{1}{\sqrt{2}}\left(e_1\pm\ii e_2\right)\;,\quad
  \theta^\pm = \frac{1}{\sqrt{2}}\left(\theta^1\mp\ii\theta^2\right)\;.
\label{eq:3}
\end{equation}
The basis vectors $e_\pm$ have spin $\pm1$. From the definition of these bases, the relations
\begin{equation}
  \theta^1(e_\pm) = \frac{1}{\sqrt{2}}  \theta^\pm(e_1)\;,\quad
  \theta^2(e_\pm) = \frac{\pm\ii}{\sqrt{2}}  \theta^\mp(e_2)
\label{eq:4}
\end{equation}
follow easily since $\theta^i(e_j)=\delta^i_j$ for $i,j = 1,2$ (or $i,j=+,-$). Let now $\mathcal{T}$ be an arbitrary rank-$r$ tensor field, then this tensor applied to $r$ copies of $e_\pm$,
\begin{equation}
  {_{\pm r}t} = \mathcal{T}(e_\pm, \ldots, e_\pm)
\label{eq:5}
\end{equation}
is a function with spin $\pm r$. Rank-$r$ tensors with defined spin properties can thus be constructed as
\begin{equation}
  \tilde T = \left({_rt}\right)\theta^+\otimes\ldots\otimes\theta^+ +
  \left({_{-r}t}\right)\theta^-\otimes\ldots\otimes\theta^-\;.
\label{eq:6}
\end{equation}

Fully symmetric tensors, which we are primarily concerned with, can be expanded into the ordered sum
\begin{equation}
  \mathcal{T} = \sum_{k=0}^r\binom{r}{k}\,\mathcal{T}_{1^{r-k}2^k}\,\underbrace{\theta^1\otimes\ldots\otimes\theta^1}_{r-k}
  \underbrace{\theta^2\otimes\ldots\otimes\theta^2}_{k}\;,
\label{eq:6a}
\end{equation}
where an exponent $k$ on an index denotes $k$ repetitions of this index. The spin-$\pm r$ functions $_{\pm r}t$ are then
\begin{equation}
  _{\pm r}t = \frac{1}{2^{r/2}}\sum_{k=0}^r\binom{r}{k}\,\mathcal{T}_{1^{r-k}2^k}\,(\pm\ii)^k\;.
\label{eq:6b}
\end{equation}
Of particular interest for us are second- and third-rank, symmetric tensors. For $r = 2$, we find
\begin{equation}
  _{\pm2}t = \frac{1}{2}\left(\mathcal{T}_{11}-\mathcal{T}_{22}\right)\pm\ii\mathcal{T}_{12}\;,
\label{eq:6c}
\end{equation}
while $r = 3$ gives
\begin{equation}
  _{\pm3}t = \frac{1}{2^{3/2}}\left[
    \left(\mathcal{T}_{111}-3\mathcal{T}_{122}\right)\pm\ii\left(3\mathcal{T}_{112}-\mathcal{T}_{222}\right)
  \right]\;.
\label{eq:6d}
\end{equation}
Quite obviously, Eqs.~(\ref{eq:6c}) and (\ref{eq:6d}) reproduce the shear and the $\mathcal{G}$-flexion, respectively, and their complex conjugates.

At any arbitrary point on the sphere, we now introduce a polar angle $\varphi$ with respect to the local $e_1$ direction and define a tangential and a radial unit vector, $e_T$ and $e_R$, respectively, by
\begin{eqnarray}
  e_R &=& R(\varphi)e_1 = \frac{1}{\sqrt{2}}R(\varphi)(e_++e_-)\;,\nonumber\\
  e_T &=& R(\varphi)e_2 = -\frac{\ii}{\sqrt{2}}R(\varphi)(e_+-e_-)\;,
\label{eq:7}
\end{eqnarray}
where $R(\varphi)$ is a matrix representing a (two-dimensional) rotation by an angle $\varphi$. Since $e_\pm$ are defined to be eigenvectors of $R(\varphi)$ with eigenvalues $\e^{\mp\varphi}$,
\begin{equation}
  e_R = \frac{1}{\sqrt{2}}\left(\e^{-\ii\varphi}e_++\e^{\ii\varphi}e_-\right)\;,\quad
  e_T = -\frac{\ii}{\sqrt{2}}\left(\e^{-\ii\varphi}e_+-\e^{\ii\varphi}e_-\right)\;,
\label{eq:8}
\end{equation}
the dual helicity basis vectors $\theta^\pm$ applied to these vectors give
\begin{equation}
  \theta^\pm(e_R) = \frac{1}{\sqrt{2}}\e^{\mp\ii\varphi}\;,\quad
  \theta^\pm(e_T) = \frac{\mp\ii}{\sqrt{2}}\e^{\mp\ii\varphi}\;.
\label{eq:9}
\end{equation}
The radial and tangential components of the spin-$\pm r$ tensor field $\tilde T$ can now be defined by the application of $\tilde T$ to $r$ copies of the radial and tangential unit vectors $e_R$ and $e_T$, respectively,
\begin{equation}
  \tilde T_R = \tilde T(e_R, \ldots, e_R)\;,\quad
  \tilde T_T = \tilde T(e_T, \ldots, e_T)\;.
\label{eq:10}
\end{equation}
With the expansion Eq.~(\ref{eq:6}), this gives
\begin{eqnarray}
  \tilde T_R &=& \frac{1}{2^{r/2}}\left[(_rt)\e^{-\ii r\varphi}+(_{-r}t)\e^{\ii r\varphi}\right]\;,\nonumber\\
  \tilde T_T &=& \frac{(-1)^r}{2^{r/2}}\left[\ii^r(_rt)\e^{-\ii r\varphi}+(-\ii)^r(_{-r}t)\e^{\ii r\varphi}\right]\;.
\label{eq:10a}
\end{eqnarray}
If we insert the expressions for $_{\pm r}t$ for symmetric tensors from Eq.~(\ref{eq:6b}) here, we finally obtain
\begin{equation}
  \tilde T_R = \frac{1}{2^{r-1}}\sum_{k=0}^r\binom{r}{k}\,\mathcal{T}_{1^{r-k}2^k}\,\Re\left(\ii^k\e^{-\ii r\varphi}\right)
\label{eq:10b}
\end{equation}
for the radial component and
\begin{equation}
  \tilde T_T = \frac{(-1)^r}{2^{r-1}}\sum_{k=0}^r\binom{r}{k}\,\mathcal{T}_{1^{r-k}2^k}\,\Re\left(\ii^{k+r}\e^{-\ii r\varphi}\right)
\label{eq:10c}
\end{equation}
for the tangential component. Quite obviously, for even rank $r$, these components are related by $\tilde T_T = \ii^r\tilde T_R$. For a symmetric, second-rank tensor, these general Eqs.~(\ref{eq:10b}) and (\ref{eq:10c}) specialise to
\begin{equation}
  \tilde T_R = \frac{1}{2}\left(\mathcal{T}_{11}-\mathcal{T}_{22}\right)\cos2\varphi+\mathcal{T}_{12}\sin2\varphi\;,\quad
  \tilde T_T = -\tilde T_R\;.
\label{eq:10d}
\end{equation}

\subsection{Application to gravitational lensing}

We now proceed to apply this general formalism to second- and third-order partial derivatives of the effective lensing potential $\psi$. With the second-order partial derivatives of the lensing potential $\psi$, we can define the rank-$2$ tensor
\begin{equation}
  \psi^{(2)} = \partial_i\partial_j\psi\,\theta^i\otimes\theta^j\;,
\label{eq:11}
\end{equation}
from which we can project out functions with spin $0$ and spins $\pm2$ by
\begin{equation}
  {_0\psi^{(2)}} = \psi^{(2)}(e_+, e_-)\;,\quad
  {_{\pm2}\psi^{(2)}} = \psi^{(2)}(e_\pm, e_\pm)\;.
\label{eq:11a}
\end{equation}
Straightforward calculation shows that the spin-$0$ function is the lensing convergence,
\begin{equation}
  {_0\psi^{(2)}} = \kappa\;,
\label{eq:11b}
\end{equation}
while the spin-$\pm2$ functions are related to the shear by
\begin{equation}
  {_2\psi^{(2)}} = \gamma = \gamma_1+\ii\gamma_2\;,\quad
  {_{-2}\psi^{(2)}} = \gamma^* = \gamma_1-\ii\gamma_2\;.
\label{eq:11c}
\end{equation}
Accordingly, we define a rank-$2$ \textit{shear tensor} $\tilde\gamma$ from the spin-$\pm2$ components of the tensor $\psi^{(2)}$ introduced in (\ref{eq:6}) above,
\begin{equation}
  \tilde\gamma = \gamma\,\theta^+\otimes\theta^+ + \gamma^*\,\theta^-\otimes\theta^-\;.
\label{eq:11d}
\end{equation}

The usual tangential and cross components of the shear, $\gamma_T$ and $\gamma_\times$, are then straightforwardly introduced as the results of the tensor $\tilde\gamma$ applied to $(e_T, e_T)$ and $(e_T, e_R)$,
\begin{equation}
  \gamma_T = \tilde\gamma(e_T, e_T) = -\Re\left(\gamma\e^{-2\ii\varphi}\right)\;,\quad
  \gamma_\times = \tilde\gamma(e_T, e_R) = \Im\left(\gamma\e^{-2\ii\varphi}\right)
\label{eq:12}
\end{equation}
which reproduce the common definitions
\begin{equation}
  \gamma_T = -\gamma_1\cos2\varphi-\gamma_2\sin2\varphi\;,\quad
  \gamma_\times = -\gamma_2\cos2\varphi + \gamma_1\sin2\varphi\;.
\label{eq:13}
\end{equation}

Applying the general formalism to the flexion now, we begin with the rank-$3$ tensor
\begin{equation}
  \psi^{(3)} = \partial_i\partial_j\partial_k\psi\,\theta^i\otimes\theta^j\otimes\theta^k\;,
\label{eq:14}
\end{equation}
and extract its spin-$1$ component by the operation
\begin{equation}
  {_1\psi^{(3)}} = \psi^{(3)}(e_+, e_+, e_-)\;.
\label{eq:15}
\end{equation}
A brief calculation shows that the result is related to the $\mathcal{F}$-flexion by
\begin{equation}
  {_1\psi^{(3)}} = \frac{\cal F}{\sqrt{2}}\;.
\label{eq:16}
\end{equation}
Since the ${\cal F}$-flexion is a vector, its radial and tangential components are trivially given by
\be
  {\cal F}_R = {\cal F}_1\cos \varphi + {\cal F}_2 \sin\varphi\;, \quad
  {\cal F}_T = {\cal F}_1\sin \varphi - {\cal F}_2 \cos\varphi\;.
\label{eq:18}
\ee
Specialising the general Eqs.~(\ref{eq:10b}) and (\ref{eq:10c}) to $r = 3$, and multiplying by a factor of two for convenience, we find the radial and the tangential components of the $\mathcal{G}$-flexion,
\begin{eqnarray}
  {\cal G}_R &=& {\cal G}_1\cos3\varphi+{\cal G}_2\sin3\varphi\;,\nonumber\\
  {\cal G}_T &=& {\cal G}_1\sin3\varphi-{\cal G}_2\cos3\varphi\;.
\label{eq:20}
\end{eqnarray}

\subsection{Singular isothermal elliptical halo}

Having laid out a general formalism for defining the tangential and radial components of a symmetric rank-$r$ tensor field on the sphere, we now proceed to calculate the radial and tangential components of the $\mathcal{G}$-flexion for a dark-matter halo modelled as a Singular Isothermal Ellipsoid (SIE) with ellipticity $\epsilon$. The halo ellipticity is defined by $\epsilon=(\theta_a-\theta_b)/(\theta_a+\theta_b)$, where $\theta_a$ and $\theta_b$ are the major and minor axes, respectively. The dimensionless surface mass density for a SIE halo is
\be
  \kappa = {\theta_{\rm E} (1-\epsilon) \over 2 \rho}\;,
\label{kappa}
\ee
where $\rho=\sqrt{f^2\theta_1^2 + \theta_2^2}$ is a radial coordinate which is constant on elliptical contours, and $f=\theta_b/\theta_a$ is the axis ratio. Then from our result (\ref{eq:20}) for the two $\mathcal{G}$-flexion components, we have
\bea
  {\cal G}_R &=& - {3\over 2} {\theta_{\rm E} (1-\epsilon) \over \theta \rho}\;,\\
  {\cal G}_T &=& - {(1-\epsilon) (1-f^2)} {\theta_{\rm E} \theta_1\theta_2 \over 2 \theta \rho^3}\;,
\label{eq:22}
\eea
where $\theta = \sqrt{\theta_1^2 + \theta_2^2}$ is the ordinary circular radius. The ratio between the tangential and the radial components of the $\mathcal{G}$-flexion is similar to that of the $\mathcal{F}$-flexion,
\be
  {{\cal G}_T\over {\cal G}_R} =
  {2 \epsilon \sin 2 \varphi \over 3 (1- 2\epsilon \cos 2 \varphi + \epsilon^2)}
  = {{\cal F}_T\over 3{\cal F}_R}\;,
\label{ratio3}
\ee
except that the ratio for the $\mathcal{G}$-flexion is one third times that for the $\mathcal{F}$-flexion. In case of the SIE halo mass distribution, ${\cal G}_R$ is three times larger than ${\cal F}_R$. However, the tangential components of both the $\mathcal{F}$- and $\mathcal{G}$-flexions are only affected by the ellipticity of lensing mass distribution, i.e.~they are equal in magnitude. Thus the ratio between the components of the $\mathcal{G}$-flexion is one third times that of the $\mathcal{F}$-flexion. Thus, if we define the ratio for the $\mathcal{G}$-flexion as $q\equiv 3({\cal G}_T/{\cal G}_R)$, an identical approach as for the $\mathcal{F}$-flexion can be employed also for the $\mathcal{G}$-flexion. In particular, the estimators $\langle q\rangle$ and $\langle q^2\rangle$ for the halo ellipticity are then the same as in \citet{2011MNRAS.417.2197E}.

Ignoring noise, the ratio $q$ for the ${\cal F}$- and ${\cal G}$-flexions will be identical. Then, the ratio for the ${\cal G}$-flexion would be redundant with the ratio from the $\mathcal{F}$-flexion and would thus not provide extra information. With noise, however, we expect two sets of data from the $\mathcal{F}$- and $\mathcal{G}$-flexions with different noise properties. In particular, since the $\mathcal{G}$-flexion causes a triangular-shaped distortion of images, its signal-to-noise ratio may be substantially higher than for the $\mathcal{F}$-flexion which causes essentially a centroid shift of the background galaxy image.

If the noise contributions to the $\mathcal{F}$- and $\mathcal{G}$-flexion measurements can be considered independent, we can simply add the signal from ${\cal F}$ and ${\cal G}$, obtain twice the amount of data and achieve a lower sample variance. In reality, as pointed out by Viola et~al.(2011), the shear needs to be known for an unbiased measurement of both types of flexion. Uncertainty in the shear may thus bias the flexion and may couple the measurement noise in the different types of flexion. The noise contributions to the two types of flexion may thus be correlated, complicating the joint estimation of halo ellipticities from both types of flexion. However, lacking any detailed knowledge on the flexion noise, a simplified approach may be justified. 

In absence of noise, ${\cal F}_T={\cal G}_T$ and $3{\cal F}_R = {\cal G}_R$. The data points will be particularly noisy if the tangential or the radial components of the ${\cal F}$ and ${\cal G}$ flexions are very different, i.e.\ if the noise dominates the measurement. Thus, we only use the simulated data if ${\cal F}_T$ and ${\cal G}_T$ as well as $3{\cal F}_R$ and ${\cal G}_R$ are not very different, e.g.\ $0.5<{\cal F}_T/{\cal G}_T<2$. This may be appropriate without detailed knowledge of the flexion noise. We do not aim at reducing the noise, but only disregard the most noisy data. The information in the most noisy data will thus not be used, reducing the amount of independent data.

\begin{figure*}
\includegraphics[width=6.5cm,height=6.cm]{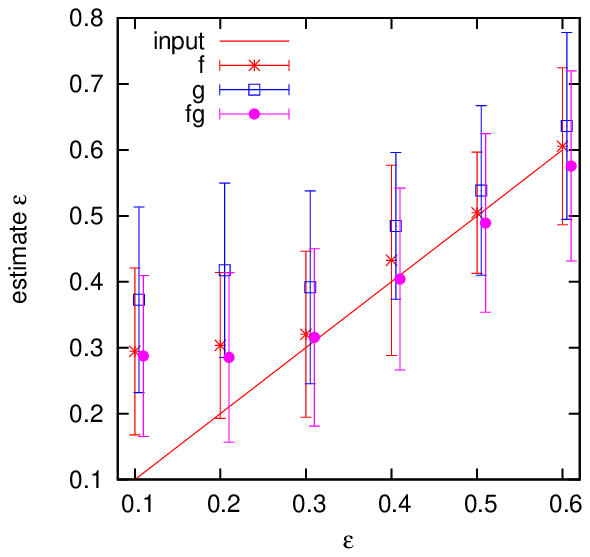}
\includegraphics[width=6.5cm,height=6.cm]{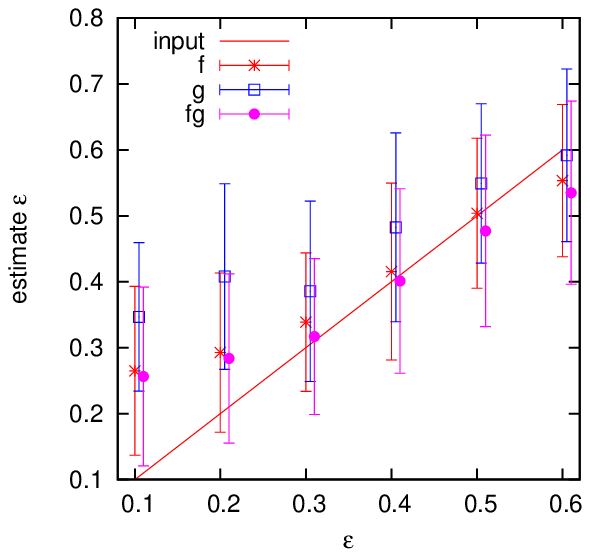}
\caption{Comparison of the ellipticity estimator with the input ellipticity (solid line) for an SIE halo. The ellipticity is estimated from 50 realizations. The error bars represent the standard deviations. In the left (right) panels, the results obtained with uncorrelated (same) noise are shown. The red stars (blue squares, purple circulars) represent the results using the $\mathcal{F}$-flexion ($\mathcal{G}$-flexion, or both).}
\label{fig:estimator}
\end{figure*}

\section{Numerical tests}

In order to test the estimator for the halo ellipticity from the $\mathcal{G}$ flexion, we use an SIE halo (Eq.~\ref{kappa}) to generate mock data. We choose a lensing halo with an Einstein radius of $\theta_{\rm E}=5''$ and place it at redshift of $z=0.4$. Background source galaxies are randomly distributed within a field sized $1'\times 1'$ behind the lens. The redshifts of the source galaxies are drawn from a Gamma distribution with $z_0=1/3$,
\be
  p(z)={z^2 \over 2z_0^3} {\rm exp}(-z/z_0)\;,
\label{redshift}
\ee
which peaks at $z=2/3$ and has a mean redshift of $\langle z\rangle =3z_0=1$.

We create mock data sets with different values for the halo ellipticity, $0.1\le\epsilon\le0.6$. For each ellipticity, we produce 50 realizations, with a spatial number density of noisy background galaxies of $40\,\mbox{arcmin}^{-2}$. The average estimate over 50 realizations is used as expected estimation, and the standard deviation is used as the errorbar. Lacking a reliable noise measurement, we adopt a simple model to include noise into our simulated data,
\begin{equation}
  {\cal G}^{\rm obs} = {\cal G}_1 + n_{g1} + \ii ({\cal G}_2 + n_{g2})\;.
\end{equation}
The noise contributions $n_{g1}$ and $n_{g2}$ are drawn from Gaussian distributions with identical variance, $\sigma_{g1} = \sigma_{g2} = 0.04\,\mbox{arcsec}^{-1} $ for each component. Synthetic data for the ${\cal F}$-flexion are also generated with Gaussian intrinsic noise characterized by the variance $\sigma_{f1} = 0.03\,\mbox{arcsec}^{-1} = \sigma_{f2}$ \citep{2007ApJ...660.1003G}.

%\bea
%{\cal F}^{\rm obs} &=& {\cal F}(1-\langle\mu|\chi^2|\rangle)+\gamma^*{\cal G};
%\nonumber\\
%{\cal G}^{\rm obs} &=& {\cal G}(1-\langle\mu|\chi^2|\rangle)+\gamma {\cal F},
%\label{corrnoise}
%\eea
%where $\langle\mu|\chi^2|\rangle$ accounts for the ellipticity
%dispersion of the galaxies. We generate $\langle\mu|\chi^2|\rangle$
%from a Gaussian distribution with variance $0.05$ (see more detail
%from Eq.~33 in Viola et~al.~2011).

Flexion data are discarded if they are taken closer to the halo center than $4''$ or farther from the halo center than $14''$, where the flexion signal will be very small and the noise is expected to dominate. Since large flexions cannot be measured \citep{2008A&A...485..363S}, data with $|{\cal G}|>1.5\,\mbox{arcsec}^{-1}$ or $|{\cal F}|>0.5\,\mbox{arcsec}^{-1}$ are also discarded. Moreover, due to the upper bound on the flexion ratio \citep{2011A&A...528A..52E}, ratios $q>3$ are discarded as well. This amounts to assuming that the halo ellipticity does not exceed $\epsilon = 0.7$. We have approximately 10 flexion data points for each realization on average. The flexion ratio $q$ is calculated for each simulated flexion measurement. The mean flexion ratio is obtained by $\bar q = (1/N)\sum_{i=1}^{N}q_i$, where $N$ is the number of images for each realization. We use the estimator
\be
  \hat \epsilon = \frac{{\rm exp}[\pi \bar q/2]-1}{{\rm exp}[\pi \bar q/2]+1}
\label{estimator}
\ee
for the halo ellipticity and ignore any redshift information in our tests.

We perform two series of tests with our estimator and show our results in Fig.~\ref{fig:estimator}. In one series, the noise of ${\cal F}$ and ${\cal G}$ in the data are generated independently (left panel). In another series of data, we use correlated noise levels, namely $n_{g1}=~(4/3)n_{f1}$ and $n_{g2}=~(4/3)n_{f2}$ (right panel). In both panels, the solid line shows the input value. The stars show the result using the $\mathcal{F}$-flexion, the open squares show the results using the $\mathcal{G}$-flexion, and the filled circles show the result using both the $\mathcal{F}$- and $\mathcal{G}$-flexions. Note that very discrepant data points are discarded as described in the previous section in the combining analysis using ${\cal F}$ and ${\cal G}$.  We can see that the estimate using the $\mathcal{G}$-flexion is significant larger than the estimate using the $\mathcal{F}$-flexion in both panels. The reason for is that the ${\cal G}$-flexion ratio signal is one third of that of ${\cal F}$-flexion. The noise thus affects the ${\cal G}$-flexion ratio more strongly. Another reason is that the noise we assume for ${\cal G}$-flexion is larger than the noise of ${\cal F}$-flexion.
In both panels, one can see that the estimations using two types of
flexion provide better results, i.e.\ lower bias, since very
noisy data are discarded. On the other hand, the error bars of
different estimations have no significant difference. In the combined
estimation using ${\cal F}$ and ${\cal G}$, since we disregard discrepant
measurements, the total number of data is not twice as that of using
only one flexion. Thus the error bars are not significantly smaller
than those using only one type of flexion.

\subsection{Centroid offset noise}

\begin{figure*}
\includegraphics[width=6.4cm,height=5.6cm]{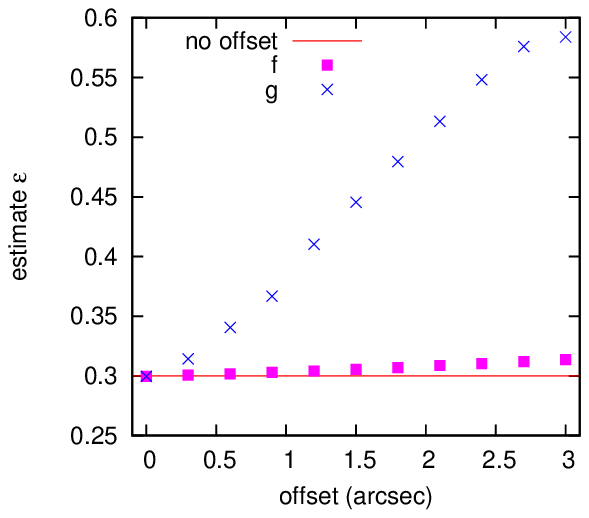}
\includegraphics[width=6.4cm,height=5.6cm]{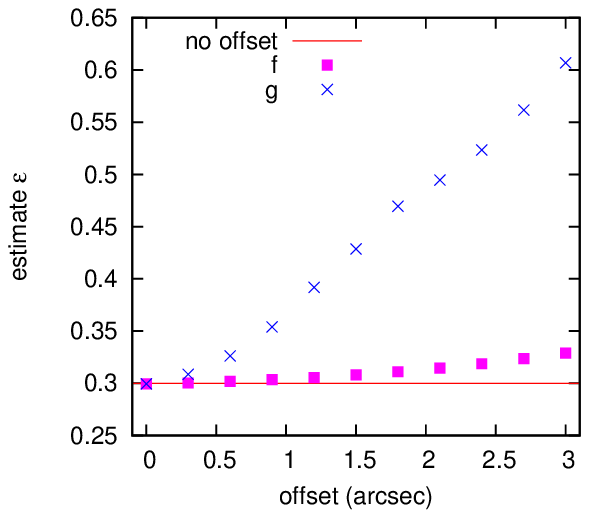}
\caption{Bias due to a centroid offset of the lens. The solid line is the input value. The left (right) panel shows the result for a centroid shift applied to the major (minor) axis only. The solid squares (crosses) are the estimates using the $\mathcal{F}$- ($\mathcal{G}$-) flexions.}
\label{fig:offset}
\end{figure*}

Since the tangential and radial flexion components are calculated with respect to the mass center of the lens, a centroid offset of the lens will affect the determination of the halo ellipticity. This bias has been studied and shown to be small using the $\mathcal{F}$-flexion \citep{2011A&A...528A..52E}. Here, we test the effect of centroid offsets on halo-ellipticity estimates from the $\mathcal{G}$-flexion. We simulate this effect by calculating the flexion ratio with respect to an assumed, perturbed centroid position $\theta_0+\delta\theta$, where $\theta_0$ is the true center of the lens and $\delta\theta$ is the offset. We apply separate and independent offsets along the major and minor axes of the lens. To isolate this effect, we generate 200 realisations of halo fields without intrinsic noise on the source galaxies for a lens with ellipticity $\epsilon=0.3$. The results are shown in Fig.~\ref{fig:offset}. The left (right) panel shows the estimate for the centroid shift applied along the major (minor) axis of the elliptical halo alone. The solid squares (crosses) present the estimate using the $\mathcal{F}$- ($\mathcal{G}$-) flexions. We can see that the bias due to the centroid offset is significant using the $\mathcal{G}$-flexion. A similar result is obtained with noisy data. Moreover, the bias due to the centroid offset dominates the effect of the intrinsic noise for the ${\cal G}$-flexion. Thus, considering the centroid offset, using the $\mathcal{G}$-flexion is more problematic than using $\mathcal{F}$-flexion, but the $\mathcal{G}$-flexion may provide a better tracer for the center of lens halo.

\begin{figure}
\includegraphics[width=6.5cm,height=6.cm]{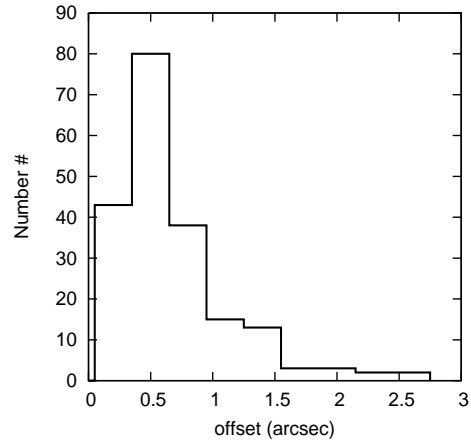}
\caption{ Histogram of the estimated halo center offset for an SIE halo with $\epsilon=0.3$ from 200 realizations.}
\label{fig:center}
\end{figure}

We can identify the center of the lens halo using the tangential components of both types of flexion, ${\cal F}_T$ and ${\cal G}_T$, since the radial flexions are less affected by the centroid offset. We thus calculate the mean tangential flexion over all the background galaxies with respect to all the candidates of lens center. Here we choose $10000$ points on a $6\times6$ arecsec$^2$ grid which is centered at the true center of the lens. The center is then determined at the position where we obtain the minimum mean tangential flexion. In Fig.~\ref{fig:center} we show our estimated halo center offset as a histogram using 200 noisy realizations for a halo with $\epsilon=0.3$. Most of our estimates of the lens center have errors smaller than one arc second. We need to point out that this is valid for a regular SIE halo without any substructures. In case of non-relaxed cluster haloes, the mass center is often ill-defined and more difficult to identify.

\section{Summary}

In this paper, we have introduced a general formalism for decomposing lensing quantities of arbitrary order and spin into components of different orientation. We have applied the formalism to the spin-3 galaxy-galaxy flexion with elliptical mass distributions. We have derived the ratio of tangential to radial flexion and given the relation to the halo ellipticity. This ratio is similar to that for the spin-1 flexion, i.e.~the ratio depends solely on the ellipticity and orientation of the lens mass distribution. Thus the spin-3 flexion can provide extra information in the estimation. The noise behavior of the two types of flexion is still unclear, i.e.~how large is the intrinsic noise and the noise correlation are for and across the ${\cal F}$ and ${\cal G}$ flexions. We can only apply a simple data selection by comparing two types of radial (tangential) flexion and discarding the very noisy data. In the numerical tests, dropping the very noisy data can reduce the bias of our estimations. However, this method reduces the amount of data and loses information. Any specific noise cross-correlation between the two types of flexion can be employed in the future noise reduction, and a smaller bias will be expected with further noise analysis.

We have found that unlike the spin-1 flexion, a misidentification of the mass centroid can introduce a significant bias in our estimates using spin-3 flexion. We thus propose to use the mean tangential flexion to identify the mass center of the lensing halo. In the test using a regular SIE halo without subhaloes, the majority of the estimated centroid offsets using our method are smaller than one arc second. In the more realistic case, e.g.~non-relaxed halo with substructures, further study is needed, possibly to be combined with other types of data, such as the X-ray surface brightness.

The shear signal can also provide constraints on the halo properties
and has a larger signal amplitude. However, shear should suffer more
from the intrinsic shape noise of the background galaxies. It can be
estimated with the simple case of a singular isothermal halo profile
with an Einstein radius $\theta_{\rm E}$. Let $\theta_{\rm E} = 1$ arc
second, for example. The shear signal at $\theta=5 \theta_{\rm E}$ is
$0.1$, while the flexion signal is $0.02/$arcsec. Since the intrinsic
standard deviation of the ellipticity is typically found to be
$\sigma_{\epsilon}\sim 0.3$, adopting a standard deviation of
$\sigma_{f}\sim 0.04/$arcsec for the flexion gives similar
signal-to-noise ratios for the shear ($1/3$) and the flexion
($1/2$). Moreover, combining shear and flexion information in the
analysis can provide tighter constraints of the halo properties
\citep{2012MNRAS.421.1443E}.

A high number density of background galaxies is crucially important. However, if the two types of flexion can be measured with sufficient accuracy, we shall already be able to place tight constraints on the halo ellipticity with current surveys, e.g.\ the HST Cosmic Evolution Survey (COSMOS) and the Canada-France-Hawaii Telescope Legacy Survey. We thus look forward to future projects with high precision images such as the Dark Energy Survey and the Euclid mission.

\section*{Acknowledgments}

We thank Shude Mao and the referee for useful comments on the
manuscript. XE thanks the Institut fuer Theoretische Astrophysik of
the University of Heidelberg for hospitality during the preparation of
this work.

\bibliographystyle{mn2e}
\bibliography{../../bib/refbooks,../../bib/lens,../../bib/flexion,../../bib/refcos,../../bib/shape}

\end{document}